# All-optically induced ultrafast photocurrents: Beyond the instantaneous coherent response

Shekhar Priyadarshi, Klaus Pierz, and Mark Bieler

*Physikalisch-Technische Bundesanstalt, 38116 Braunschweig, Germany*

It is demonstrated that the non-instantaneous response of the optically induced coherent polarization tremendously influences the real-space shift of electronic charges in semiconductors. The possibility to coherently control this real-space shift with temporally non-overlapping excitation pulses allows for the observation of a new type of shift current, which only exists for certain polarization-shaped excitation pulses and vanishes in the continuous-wave limit. In contrast to previously studied shift currents, the new current requires a phase mismatch between two orthogonal transition dipole moments and leads, within a nonlinear second-order description, to a tensor which is antisymmetric with respect to the order of the two exciting electric field amplitudes. These observations, which can even be made at room temperature and are expected to occur in a variety of semiconductor crystal classes, contribute to a better understanding of light-matter interaction involving degenerate bands. Thus, they are expected to prove important for future studies of coherent and nonlinear optical effects in semiconductors.

Controlling microscopic particles with a speed that is beyond the limits of mechanical handling has been the essence of many scientific fields for centuries.[1-3] Electronic motion now can be controlled with a femtosecond or even sub-femtosecond precision using ultrashort laser pulses. Among other applications, such coherent control schemes have been shown to result in all-optically induced currents which can be linked to nonlinear optical processes.[4-13] For the lowest-order nonlinear process, i.e., second-order in electric field, the currents resulting from above-bandgap excitation can be classified into two types: (i) injection current or circular photogalvanic current[4, 8, 14, 15] and (ii) shift current or linear photogalvanic current.[4, 5, 9] The injection current results from quantum interference of absorption processes associated with orthogonal polarization components of circularly-polarized excitation. In 18 of the 21 non-centrosymmetric crystal classes such an interference process is capable of creating an asymmetric carrier distribution in momentum space, which, in turn, leads to a current.[16] In contrast, the shift current is induced by linearly-polarized optical excitation and results from spatially displaced valance- and conduction-band wavefunctions which cause a real-space shift of the center of the electron charge during excitation. The shift current is allowed to occur in all non-centrosymmetric crystal classes.[16]

Here we employ the non-instantaneous response of the optically induced coherent polarization to coherently control the real-space shift of electronic charges with temporally non-overlapping excitation pulses. Such coherent control enables us to observe a new type of shift current which only occurs if the linear polarization of the excitation pulse rotates versus time. For pulsed excitation with a non-rotating linear polarization and in the continuous-wave limit this current is zero. Macroscopically, the new current can be linked to an antisymmetric tensor and, thus, differs from previously known symmetric shift current tensors. Our work gives important guidelines how to experimentally distinguish between symmetric and antisymmetric current response tensors. We note that the possible occurrence of photocurrents that vanish in the continuous-wave limit has been touched in a theoretical work by Nastos and Sipe considering dispersive current tensors.[17] Here, we present a simplified version of the detailed theory put forward by Sipe and coworkers[4, 16, 17] to demonstrate the underlying physics of our observations. This simple model well reproduces the experimental results and allows us to explain the occurrence of the antisymmetric shift current (ASC) with a phase mismatch between two orthogonal transition dipole moments.

The experiments are performed on exciton resonances of two nominally symmetric and unstrained (110)-oriented GaAs/Al$_{0.3}$Ga$_{0.7}$As quantum well (QW) samples comprising 40 wells with a width of 5 nm and 8 nm. The (110) orientation simplifies the experiment since it allows for generation of in-plane currents for normally incident excitation; however other orientations and semiconductors are expected to enable the same observations. The optical excitation is obtained from a pair of unchirped, time-delayed, and orthogonally polarized femtosecond laser pulses (76 MHz repetition rate, 140 fs and 5.2 nm temporal and spectral width, respectively). Their center wavelength is tuned to the n=1 light-hole exciton (lhX) energies of the samples.[18, 19] The peak intensity of each individual beam is kept at 3.5 MW/cm$^2$ (pump energy ~ 1.6 nJ), except as explicitly mentioned below, resulting in a carrier density of ~1×10$^{10}$/cm$^2$ per pulse. The electric field of the two excitation beams are aligned along the x = [001] and y = [1$\bar{1}$0] in-plane crystallographic directions of the QW samples. Varying the phase delay ($\varphi$) between the two electric fields leads to a superposition of shift and injection currents along the y direction. The shift and injection currents are maximized for $\varphi$ being equal to even (linearly polarized light, cosine dependence on $\varphi$) and odd (circularly polarized light, sine dependence on $\varphi$) multiples of $\pi/2$, respectively.[20] In a nonlinear second-order picture the current generation in this excitation geometry relies on the $\sigma^{yxy}$ and $\sigma^{yyx}$ shift current and $\eta^{yxy}$ and $\eta^{yyx}$ injection current tensor elements. These $\sigma$ and $\eta$ tensor elements are symmetric and antisymmetric, respectively, with respect to the last two Cartesian indices.[4] The currents are detected by measuring the simultaneously emitted terahertz (THz) radiation. To this end a probe beam, which can be time delayed to the excitation beams, is used for time-equivalent sampling of the THz time traces employing a typical electro-optic (EO) detection scheme.[18, 21]

In our experiments we employ the fact that the THz traces of shift and injection currents obtained from excitation of the lhX differ from each other. This allows us to suppress the contribution of the injection current to the measured THz field by proper adjustment of the delay of the probe-beam, see inset of Fig. 1. This probe-delay adjustment has been made for temporally overlapping excitation pulses. In subsequent measurements we have varied the time delay $\tau$ between the two excitation pulses and recorded the THz signal versus $\tau$. Such measurements yield THz interferograms (TIs) which oscillate with the optical frequency. During the measurement of the TIs, we have simultaneously adjusted the delay of the probe beam by $\tau/2$ to account for the change of $\tau$. We will show further below that the adopted variation of the probe beam is important, since it provides even (odd) TIs with respect to $\tau$ for symmetric (antisymmetric) tensor elements. Moreover, for an exact calibration of the phase delay φ between the two excitation fields we have simultaneously measured their electric-field correlation (EFC) using an ellipsometric technique.[21] For the discussion of the experiments we employ the phase differences between the TI and the EFC. These phase differences are equal to even and odd multiples of π/2 for a scenario in which the current response is governed by a cosine (shift current) and sine (injection current) dependence on φ, respectively. As will be seen below non-zero phase differences being equal to even multiples of π/2 will be important; these phase differences correspond to an optical excitation pulse whose linear polarization rotates slowly in time.

We now comment on the experimental results. In Fig. 1a the THz signal obtained from excitation of the n=1 lhX of the 5 nm QW at room temperature is plotted versus $\tau$. The TI shows two sharp waists on either side. Even more interesting, the interferogram is clearly asymmetric with respect to $\tau$. As shown in Fig. 1b the phase difference changes from 0 to π and -π at the position of the waists. In contrast, the TI resulting from a bulk (110)-oriented GaAs reference sample, shown in Fig. 1c, shows neither such waists nor an asymmetry and the phase difference is zero for all $\tau$. Moreover, the envelopes of the TI obtained from the bulk sample and the EFC are nearly identical, while the envelope of the TI obtained from the QW sample is considerably broader.

From Fig. 1 we can draw some important conclusions. First, the measured phase difference being equal to even multiples of π/2 proves that the contribution of injection currents to the THz signal obtained from the QW sample is negligibly small. (In the bulk sample the injection current is symmetry forbidden.) Second, the slower decay of the QW TI strongly suggests that the shift current generation is influenced by a non-instantaneous process, presumably by the optically induced coherent polarization. If the coherent polarization decays slower than the electric field of the first excitation pulse it might broaden the TI. In any case, this observation proves the possibility to coherently control the real-space shift of electrons with temporally non-overlapping excitation pulses. Interestingly we observe such an effect for the excitation of excitons in our GaAs QWs even at room temperature. The influence of dephasing on the shape of the TI is further corroborated by an intensity dependence measurement for excitation of the lhX in the 5 nm QW shown in Fig. 2a and by a temperature dependence measurement for excitation of the lhX in the 8 nm QW shown in Fig. 2b. The plots show a decrease of asymmetry as well as a faster decay of the TI envelope for an increase of optical excitation intensity or temperature. Additionally, Fig. 2b verifies the sample independence of the observations.

In the following we explain the peculiarities of the measured TI and start with the occurrence of the waists and the phase behavior shown in Figs. 1a and 1b, respectively. The total current contributing to the measured THz signal consists of shift currents induced in the QW region and in the bulk substrate. The shift current obtained from excitation of the lhX in the QW region is oppositely oriented as compared to the shift current obtained from excitation of continuum carriers in the bulk substrate (where heavy-hole carriers dominate) and in the QW region (where the onset of the heavy-hole continuum overlaps with the lhX).[9] For overlapping excitation pulses, i.e., near τ=0, the shift current resulting from excitation of such continuum carriers dominates the current response, as only ~35% of the incident optical power is absorbed in the QW region. However, the coherent polarization obtained from excitation of the lhX in the QW region dephases considerably slower than the one obtained from excitation of the heavy-hole continuum in the bulk substrate and the QW region. Therefore, for certain $\tau$ the shift current resulting from excitation of the lhX becomes as large as the shift current resulting from excitation of continuum carriers and leads to a nearly perfect waist in the TI. For larger time delays the lhX shift current even dominates the current response. This current reversal produces a phase jump of π at the position of the waists, see Fig. 1b.

While the interference of shift currents resulting from the lhX and continuum carriers can be used to explain the waists in the interferogram, it cannot explain its asymmetry. To theoretically analyze this effect we adopt a model based on the Bloch equations in the length gauge as introduced by Sipe et al.[4] but restrict our analysis to the time domain. Equations for injection $J_{IC}$ and shift $J_{SC}$ currents can be obtained by taking the trace of a generalized velocity matrix $v_{nn}^y \delta_{nm} - \frac{e}{\hbar} r_{nm;y} \cdot E$ times the density matrix $\rho$ and can be written for currents along the y direction as: $J_{IC}^y(t) = e \int \frac{d\mathbf{k}}{8\pi^3} \sum_n v_{nn,\mathbf{k}}^y \rho_{nn,\mathbf{k}}^{(2)}$ and $J_{SC}^y(t) = -\frac{e^2}{\hbar} \int \frac{d\mathbf{k}}{8\pi^3} \sum_{n,m} \rho_{nm,\mathbf{k}}^{(1)} r_{mn;y} \cdot E$, respectively. Here, $|n\rangle$ and $|m\rangle$ are the two bands involved in the excitation process, $v_{nn}^y$ is the group velocity of the n$^{th}$ band, $\boldsymbol{r}_{nm}$ is the interband transition dipole moment, and $\boldsymbol{r}_{mn;y} = \frac{\partial \boldsymbol{r}_{mn}}{\partial k^y} - i(\xi_{mm} - \xi_{nn})\boldsymbol{r}_{mn}$ is the generalized derivative of the transition dipole moment.[4] It includes the so called connections $\xi_{mm}$ and $\xi_{nn}$ and denotes the electronic shift during transition from band $|m\rangle$ to $|n\rangle$. It is important to mention that the two terms in $r_{mn;y}$ ensure its phase invariance and both contribute to the shift along y. Neither of the two terms alone can be attributed to a certain physical process.[22] The optical electric field is denoted by $\boldsymbol{E}(t) = \boldsymbol{E}_{env}(t)\exp(-i\omega_c t) + c.c.$ is with $\boldsymbol{E}_{env}(t)$ and $\omega_c$ being the real electric field envelope and the center frequency, respectively. From the above equations it is easy to show that the injection current results from an asymmetrically distributed population in momentum space while the shift current results from a real-space electronic shift. In following analysis, we concentrate on shift currents only, since the contribution of injection currents to our measured THz signal are negligibly small.

After obtaining an expression for the density matrix element $\rho^{(1)}_{nm,\mathbf{k}}$ from first-order perturbation solution of the Bloch equations for the case of resonant excitation and providing a time-delayed pulse pair as excitation $\boldsymbol{E}_{env}(t) = \hat{x}E^x_{env}(t-\tau)\exp(i\omega_c\tau) + \hat{y}E^y_{env}(t)$ it is straightforward to derive an equation for the shift current which can be split into two different terms:

$$J^y_{SSC}(t) = \int \frac{d\mathbf{k}}{8\pi^3}\left[\sigma_s^{yxy} f^x_{env}(t-\tau)E^y_{env}(t) + \sigma_s^{yyx} f^y_{env}(t)E^x_{env}(t-\tau)\right]\cos(\omega_c\tau), \qquad (1)$$

$$J^y_{ASC}(t) = \int \frac{d\mathbf{k}}{8\pi^3}\left[\sigma_a^{yxy} f^x_{env}(t-\tau)E^y_{env}(t) + \sigma_a^{yyx} f^y_{env}(t)E^x_{env}(t-\tau)\right]\cos(\omega_c\tau), \qquad (2)$$

where we define $\sigma_s^{yxy} = \frac{2e^3}{\hbar^2}Im(r^x_{12}r^y_{21;y} - r^y_{21}r^x_{12;y}) = \sigma_s^{yyx}$ and $\sigma_a^{yxy} = \frac{2e^3}{\hbar^2}Im\left(\frac{\partial r^x_{12}r^y_{21}}{\partial k^y}\right) = -\sigma_a^{yyx}$. The non-instantaneous response of the coherent polarization is expressed through a modified electric-field envelope $f^{x,y}_{env}(t) = \exp(-\gamma t)\int_{-\infty}^{t} E^{x,y}_{env}(x)\exp(\gamma x)\,dx = E^{x,y}_{env}(t)\gamma^{-1} - \dot{E}^{x,y}_{env}(t)\gamma^{-2} + \ddot{E}^{x,y}_{env}(t)\gamma^{-3} - \cdots$, with $\gamma$ being the dephasing rate and the series being obtained from integrations by parts. The current component $J^y_{SSC}(t)$ corresponds to the previously known shift current. Its strength is maximized for $\tau = 0$ and governed by the tensor $\sigma_s$, which is symmetric in the last two Cartesian indices. Thus, in the following we refer to it as symmetric shift current (SSC). However, there is another current component $J^y_{ASC}(t)$ whose tensor element $\sigma_a$, is antisymmetric. This current is zero for $\tau = 0$ and maximized for positive and negative temporal delays between the two excitation pulses determined by higher-order terms of $f^{x,y}_{env}(t)$. For $t = \tau/2 + c$, with $c$ being an arbitrary constant, it is easy to show that $J^y_{SSC}(\tau)$ and $J^y_{ASC}(\tau)$ are even and odd functions of $\tau$, respectively. This is shown in Fig. 3a, where we have plotted the envelopes of $J^y_{SSC}(\tau)$ and $J^y_{ASC}(\tau)$. It is important to note that the same behavior is obtained for the corresponding TIs, independent of the transfer function of the EO detection. For very fast dephasing all higher-order terms in the modified electric-field envelope vanish, such that $J_{ASC}^y(t) = 0$ and only $J_{SSC}^y(t)$ survives. Obviously, the same applies for the continuous-wave limit. This visualizes the importance of the non-instantaneous response of the coherent polarization for the observation of the ASC and shows a remarkable difference between $J_{ASC}$ and $J_{SSC}$ even though both result from a real-space shift of charges. The reason why both a symmetric and antisymmetric tensor are obtained for the shift current but not for the injection current (the injection current tensor is purely antisymmetric with a $\sin(\omega_c \tau)$ dependence) is due to the generalized derivative of the transition dipole moment, which only appears in the equation for the shift current. Our analysis also directly proves the importance of using effective Hamiltonians for degenerate band in an electric field, which are invariant to a phase transformation, see Ref. [23] and references therein. For a further insight on the ASC we take $\boldsymbol{r}_{mn} = |\boldsymbol{r}_{mn}| e^{-i\theta_{mn}}$ and directly obtain $\sigma_a^{yxy} = Im\left(\frac{\partial}{\partial k^y}|r_{12}^x||r_{12}^y| e^{-i(\theta_{12}^x - \theta_{12}^y)}\right)$ showing that the ASC requires a phase mismatch of orthogonal transition dipole moments. This is in contrast to the SSC, which persists for all phase relationships between the transition dipole moments. Moreover, as usual for antisymmetric tensors no ASC is obtained if the two excitation fields are polarized along the same direction. Due to the necessary involvement of an antisymmetric tensor, the ASC is expected to occur in 18 of the 21 non-centrosymmetric crystal classes.[4]

For a simulation of the experimentally obtained TI shown in Fig. 1a we apply the above equations to a two-level model and, thus, neglect any *k* dependence. Considering SSC from the QW region and the bulk substrate and ASC from the QW region and applying an additional transfer function of the EO detection,[24] we obtain an excellent agreement with the experiment, see Fig. 3b, and estimate the ratio $\sigma_a / \sigma_s \approx -0.14$ for the lhX of the 5 nm QW. With respect to the omission of injection currents in the simulations, it should be noted that for resonant excitation, i.e., no detuning, shift and injection currents are orthogonal functions with respect to $\tau$. Thus, the injection current cannot interfere with the shift current and cannot cause an asymmetric envelope of the TI. Since this asymmetry can only be explained by considering the ASC whose envelope is an odd function of $\tau$, we believe that our experiments undoubtedly confirm the existence of the ASC. A previous theoretical study based on spectrally dispersive current tensors, which are equivalent to temporally non-instantaneous polarization dynamics, has already mentioned the possible occurrence of currents that vanish in the continuous-wave limit.[17] On the other hand, the experimental distinction between different current components induced by ultrashort optical pulses has also been considered to be extremely difficult.[16] Thus we believe that our work not only contributes to the understanding of general aspects of light-matter interaction but also gives important guidelines how to distinguish between the symmetric and anti-symmetric part of asymmetric response tensors.

In conclusion, we have presented evidence for the existence of a new all-optically induced current which is observable only for non-instantaneous dephasing of the optically induced polarization. Although the current stems from a real-space shift of carriers, it is significantly different than previously known shift currents, since the underlying tensors are antisymmetric with respect to the two electric excitation fields. The fact that the current is expected to occur in a variety of semiconductors and the clear detection even at room temperature in our (110)-oriented GaAs QWs shows great promise for novel coherent applications. In particular this includes the possibility for simultaneous coherent control of commutative and non-commutative operations with respect to two time-delayed pulses. Our work might also affect novel developments of optical-to-electrical converters for ultrafast (THz) communications using time-varying optical polarization as another degree of freedom.

The authors thank Holger Marx for technical assistance and the Deutsche Forschungsgemeinschaft (DFG) for financial support.

**Figure 1**

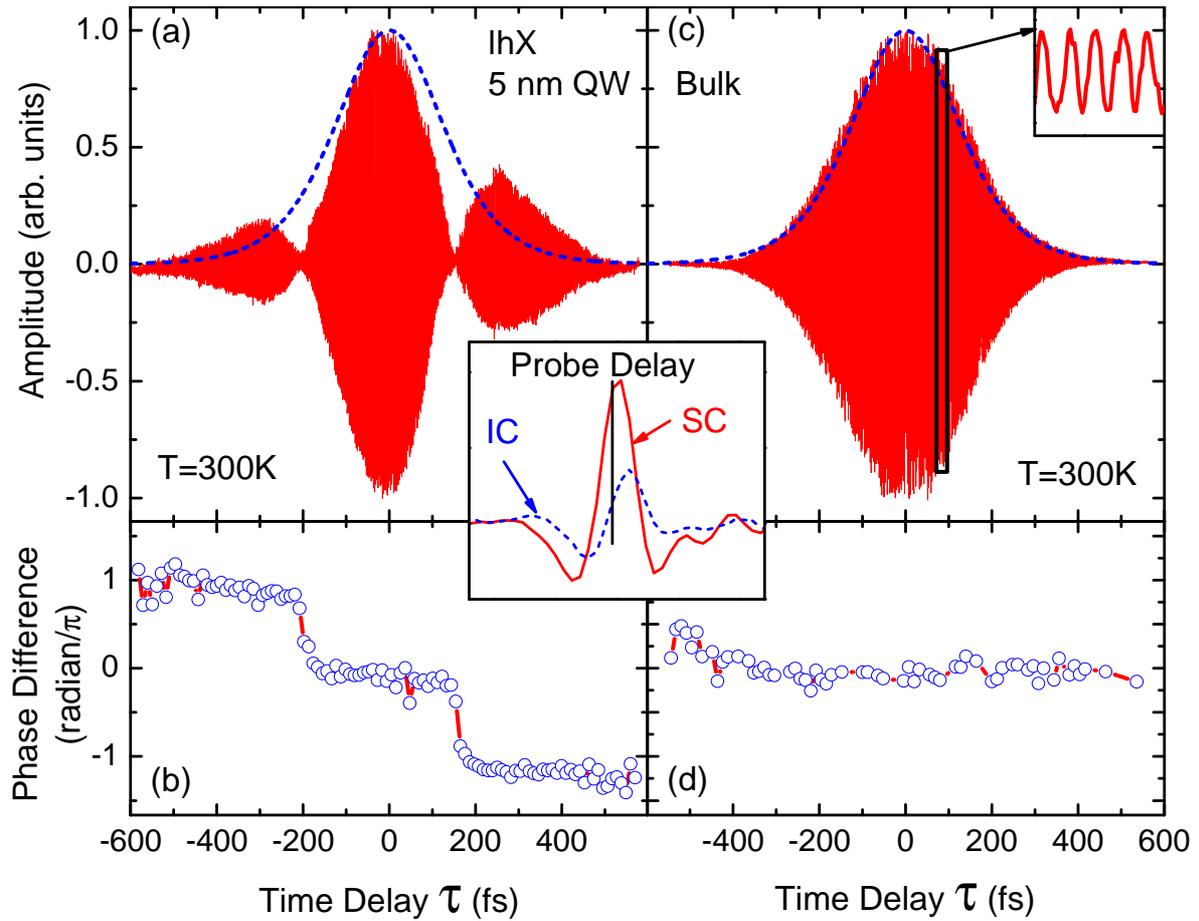

Fig. 1 (a) THz interferogram obtained from excitation of the lhX in the 5 nm QW at room temperature. The envelope of the electric-field correlation is shown as dashed line. Inset: Temporal THz traces of shift (SC) and injection (IC) currents for excitation of the lhX. The vertical solid line shows the temporal position of the probe beam used to obtain the THz interferogram and this position has been varied by $\tau/2$ to account for a change in $\tau$. (b) Phase difference between the THz interferogram of the QW sample and the electric-field correlation. (c) THz interferogram obtained from excitation of the bulk sample for the same excitation conditions as in (a). The inset shows a small part of the oscillations. (d) Phase difference between the THz interferogram of the bulk sample and the electric-field correlation.

**Figure 2**

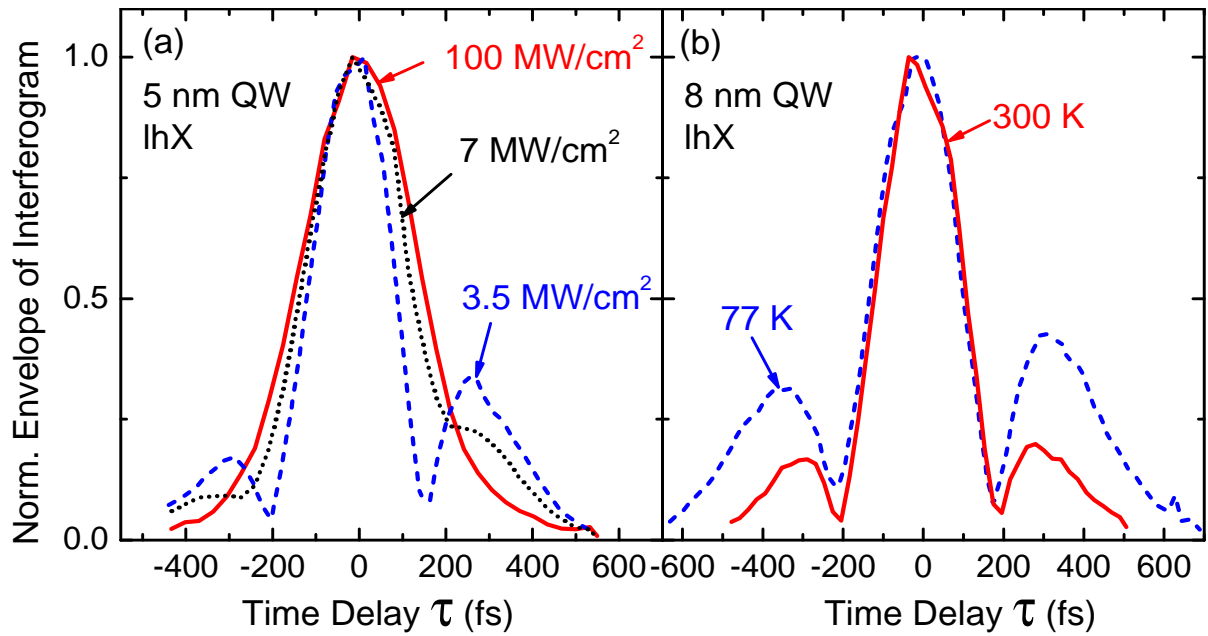

Fig. 2 (a) Intensity dependence of the envelope of the THz interferogram for excitation of the lhX in the 5 nm QW. (b) Temperature dependence of the envelope of the THz interferogram for lhX excitation of the 8 nm QW.

**Figure 3**

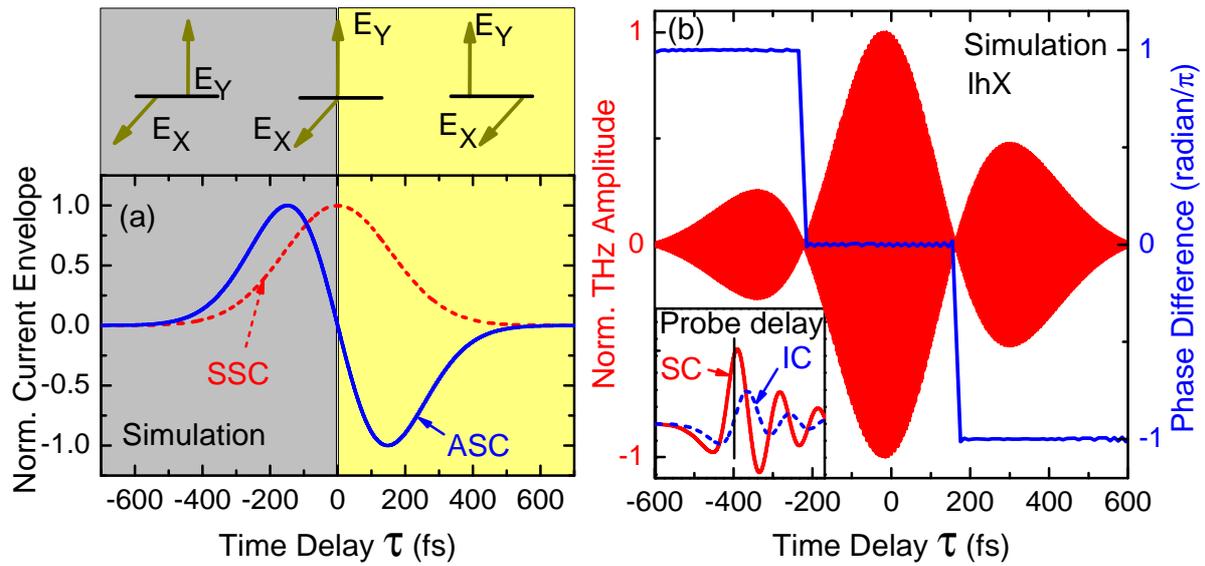

Fig. 3 (a) Envelopes of the symmetric (dashed line) and antisymmetric (solid line) shift current versus time delay $\tau$ as obtained from our simple model. (b) Simulation of the measured THz interferogram and phase difference shown in Figs. 1a and 1b. Inset: Calculation of the temporal THz traces resulting from shift (SC) and injection (IC) currents for excitation of the lhX in the QW sample. The vertical solid line shows the temporal position at which the THz signal has been taken to construct the THz interferogram and this position has been varied by $\tau/2$ to account for a change in $\tau$.